\documentclass[pra,aps,superscriptaddress,amssymb,amsmath,reprint,noeprint,floatfix]{revtex4-2}
\usepackage{natbib}
\usepackage{braket}
\usepackage{bm}
\usepackage[normalem]{ulem}
\usepackage{siunitx}

\usepackage{graphicx}
\usepackage[colorlinks=true,allcolors=blue]{hyperref}%
\usepackage{xcolor}
\hypersetup{bookmarksnumbered}
\newcommand{\uvec}[1]{\boldsymbol{\hat{\mathbf{#1}}}}

\begin{document}

\title{Optical force and torque on a spinning dielectric sphere}
\date{\today}

\author{Hengzhi Li}
\affiliation{Department of Physics, City University of Hong Kong, Tat Chee Avenue, Kowloon, Hong Kong, China}
\author{Wanyue Xiao}
\affiliation{Department of Physics, City University of Hong Kong, Tat Chee Avenue, Kowloon, Hong Kong, China}
\author{Tong Fu}
\affiliation{Department of Physics, City University of Hong Kong, Tat Chee Avenue, Kowloon, Hong Kong, China}
\author{Zheng Yang}
\affiliation{Department of Physics, City University of Hong Kong, Tat Chee Avenue, Kowloon, Hong Kong, China}
\author{Shubo Wang}\email{shubwang@cityu.edu.hk}
\affiliation{Department of Physics, City University of Hong Kong, Tat Chee Avenue, Kowloon, Hong Kong, China}

\begin{abstract}
 Optical force can enable precise manipulations of small particles for various applications. It is well known that an isotropic lossless dielectric sphere is only subject to forward optical force under the illumination of an electromagnetic plane wave. By using rigorous full-wave simulations, we show that such a sphere can experience a lateral optical force and an optical torque besides the conventional longitudinal force, if it spins with a constant angular velocity. The emergence of the unusual optical force and torque is attributed to the breaking of mirror and cylindrical symmetries by the spinning motion. Using the multipole expansion in source representation, we illustrate how the spinning-induced effective bi-anisotropy generates the lateral force and torque on the sphere through the interference of electric and magnetic multipoles. We also uncover the effect of Sagnac frequency splitting on the optical force and torque. The results contribute to the understanding of the optical force and torque in moving media and can be applied to realize unconventional optical manipulations of small particles.
\end{abstract}
\maketitle

\section{\label{sec: I. Introduction}Introduction}
Light can exert force and torque on nanoparticles according to the momentum conservation. Circularly polarized light carries both linear momentum and spin angular momentum, which can be changed when light is scattered or absorbed by particles, hence inducing optical force and torque on the particles. Ever since Arthur Askin realized laser beam trapping of small particles \cite{Ashkin1970} , optical manipulation using optical force and torque has developed expeditiously in theory and applications and has played an indispensable role in physics, chemistry, and biology \cite{Riccardi2023,Xin2020,Jia2023,Bustamante2021}. In recent years, unconventional optical forces and torques have gained increasing attention for their intriguing physics and remarkable applications \cite{Chen2011,Wang2014,Wang2016,Zhu2018,Zhu2020,Toftul2023,xu2024gradient}. The emergence of these unconventional forces and torques are usually attributed to the subtle symmetry breaking induced by the structures or optical fields. 


One intriguing type of unconventional optical forces is the lateral optical force, i.e., optical force directed perpendicularly to the linear momentum of incident light \cite{Shi2023}. Lateral optical force usually appears in a system with symmetry breaking in the lateral direction. Such lateral asymmetries may originate from structural geometry or engineered properties of electromagnetic waves. In addition to the straightforward morphology asymmetry of scatterers, the presence of a substrate and longitudinal spin can induce asymmetric scattering in the lateral directions and give rise to lateral optical forces \cite{Wang2014,rodriguez2015lateral}. It was shown that transverse spin angular momentum  can exert a lateral force on chiral particles \cite{Hayat2015}, and transverse spin momentum can exert a lateral force on dielectric particles \cite{Bliokh2014}. Recent advances also demonstrate substrate-free lateral forces exerted on geometrically symmetric scatterers \cite{Chen2020,Shi2022,Nan2023}. These lateral forces are realized by exploiting the hidden symmetry breaking in the lateral directions, which can be achieved with multi-wave interference \cite{Shi2022}, duality symmetry breaking \cite{Chen2020}, or the interplay of multipoles \cite{Nan2023}. Exploring new mechanisms of substrate-free lateral optical forces can facilitate high-efficiency optical manipulation of small particles.


Optical torque can be induced on small particles with material absorption or broken cylindrical symmetry. An electromagnetic plane wave cannot induce optical torque on a cylindrically-symmetric particle made of an isotropic and homogeneous material without loss, since the particle cannot change the total optical angular momentum via scattering. If material loss is introduced into the particle, the plane wave can induce an optical torque with the magnitude proportional to the particle’s absorption cross section \cite{jackson2021classical,Chaumet2007}. Light can also induce optical torque on anisotropic particles, where the anisotropy may originate from the materials (e.g., birefringent materials) or the geometric shapes (e.g, ellipsoid) \cite{Higurashi1999,Hoang2016,Ahn2018,chen2018,shi2022inverse}. For small anisotropic particles, the cylindrical asymmetry can result in the excitation of an electric dipole $\bold{p}$ that is not parallel or anti-parallel to the electric field $\bold{E}$. Consequently, an optical torque proportional to $\text{Re} [\bold{p}\times \bold{E}^*]$ can be induced on the particle \cite{Riccardi2023}. Besides, nonlinear light scattering can also be utilized to generate optical torque \cite{Toftul2023}. The above conventional optical torques rely on absorption or breaking the cylindrical symmetry with some static properties of the particles. Optical torque can also arise from the symmetry breaking induced by the dynamic motion of structures, which can exhibit unusual properties that cannot be realized in conventional stationary structures \cite{Movassagh2013} . Moving media belong to the general time-varying systems that have attracted growing attention recently \cite{Galiffi2022,Yin2022}. Optical systems with moving media can break the time-reversal symmetry and enable nontrivial light manipulations, such as optical isolation and nonreciprocal wavefront modulations \cite{Huang2018,Maayani2018,Shi2021,yang2024nonreciprocal}. 


In this article, we present a theoretical study of the optical force and torque induced by an electromagnetic plane wave on a dielectric sphere under spinning motion. The spinning sphere exhibits effective bi-anisotropic properties that are fundamentally decided by the relativistic constitutive relations. We show that, under the illumination of a simple plane wave, the spinning sphere is subject to unusual optical forces and torques that do not exist in stationary systems, including lateral optical force and optical torque. These optical forces and torques are closely related to the breaking of mirror and cylindrical symmetries by the spinning motion of the sphere. We also find that the Sagnac frequency splitting can strongly influence the optical forces and torques, leading to dramatic variations in their spectra. We apply multipole expansions to understand the phenomena based on the interference of the electric and magnetic multipoles induced in the sphere. 

The article is organized as follows. In Sec. II, we introduce the methodology for simulating a spinning sphere and calculating the optical force and torque induced on the sphere. We also introduce the analytical multipole expansion method in source representation used to understand the origin of the force and torque. In Sec. III, we present the numerical and analytical results of the optical force and torque induced on a spinning sphere by plane waves. We will focus on two scenarios: the plane wave incident direction is parallel (in III.A) and perpendicular (in III.B) to the spinning axis of the sphere. We discuss their unusual properties and explain the associated phenomena with the multipole expansion. We draw the conclusion in Sec. IV.    

\section{\label{sec: II. MST,SourceRep}	METHODOLOGY }

Moving media exhibit bi-anisotropic electromagnetic properties due to special relativity \cite{Minkowski1910,sommerfeld2013electrodynamics}. The constitutive relations for an axially symmetric object made of isotropic material in rotation motion can be expressed as 
\cite{Movassagh2013,Ridgely1998,Ridgely1999}
\begin{equation}
\begin{aligned}
   & \mathbf{D}+\mathbf{v} \times \frac{\mathbf{H}}{c^{2}}=\varepsilon(\mathbf{E}+\mathbf{v} \times \mathbf{B}),\\
   & \mathbf{B}+\mathbf{E} \times \frac{\mathbf{v}}{c^{2}}=\mu(\mathbf{H}+\mathbf{D} \times \mathbf{v}),
\label{eqn:1}
\end{aligned}
\end{equation}
where $\varepsilon$ and $\mu$ are the absolute permittivity and permeability of the material at rest; $\bold{v}$ is the linear velocity of a point on the object. Since $\bold{v}$ depends on the relative position in the object, the above equation indicates that the spinning object is effectively inhomogeneous and bianisotropic. For a spinning sphere, Eq. (\ref{eqn:1}) can be formulated into a matrix form
\begin{equation}
\left[\begin{array}{l}
\mathbf{D} \\
\mathbf{B}
\end{array}\right]=\left[\begin{array}{cc}
\bar{\varepsilon} & \bar{\chi}_{\mathrm{em}} \\
\bar{\chi}_{\mathrm{me}} & \bar{\mu}
\end{array}\right]\left[\begin{array}{l}
\mathbf{E} \\
\mathbf{H}
\end{array}\right]
\label{eqn:2}
\end{equation}
where $\bar{\varepsilon}$, $\bar{\mu}$, $\bar{\chi}_{\mathrm{me}}$, and $\bar{\chi}_{\mathrm{em}}$ are rank two tensors that characterize the effective material properties of the spinning sphere. They have the following elements in spherical coordinates system 
\begin{equation}
\left[\begin{array}{cc}
\bar{\varepsilon} & \bar{\chi}_{\mathrm{em}} \\
\bar{\chi}_{\mathrm{me}} & \bar{\mu}
\end{array}\right]_{r \theta \phi}=\frac{1}{\beta} \\
\left[\begin{array}{cccccc}
\alpha \varepsilon & 0 & 0 & 0 &  \gamma  & 0 \\
0 & \alpha\varepsilon & 0 & -\gamma  & 0 & 0 \\
0 & 0 & \beta \varepsilon & 0 & 0 & 0 \\
0 & -\gamma  & 0 & \alpha \mu & 0 & 0 \\
\gamma  & 0 & 0 & 0 & \alpha \mu & 0 \\
0 & 0 & 0 & 0 & 0 & \beta \mu
\end{array}\right]
\label{eqn:3}
\end{equation}
Here, $\alpha=c^{2}-v^{2}$, $\beta=c^{2}\left(1-\epsilon \mu v^{2}\right)$, $\gamma=(1-\varepsilon\mu c^{2})v$. $c$ is the speed of light in vacuum, $\epsilon$, $\mu$ are absolute permittivity and permeability of the material. $v=\varOmega r \text{sin}\theta$ with $\varOmega$ and $r$ being the angular velocity and radius of the sphere. The above constitutive relations can be implemented in common numerical simulation packages, such as COMSOL Multiphysics based on the finite-element method. Then, the total electric and magnetic fields of the spinning sphere under external excitation can be obtained by solving the Maxwell equations under the open boundary condition.

The optical force $\mathbf{F}$ and torque $\mathbf{G}$ exerted on the sphere can be calculated with the Maxwell stress tensor (MST) approach as \cite{jackson2021classical}
\begin{equation}
\mathbf{F}=\oint_{S}\langle\overline{\mathbf{T}}\rangle \cdot \hat{\mathbf{n}} da,
\label{eqn:4}
\end{equation}
\begin{equation}
\mathbf{G}=\oint_{S} \mathbf{r} \times\langle\overline{\mathbf{T}}\rangle \cdot \hat{\mathbf{n}} da,
\label{eqn:5}
\end{equation}
where $S$ is an arbitrary closed surface enclosing the spinning sphere and $\uvec{n}$ is the outward unit normal vector on the surface. Here, $ \langle\overline{\mathbf{T}}\rangle $ is the time-averaged Maxwell stress tensor with the following elements
\begin{equation}
\left\langle T_{i j}\right\rangle=\frac{1}{2} \operatorname{Re}\left[\varepsilon_{0} E_{i} E_{j}^{*}-\frac{1}{\mu_{0}} B_{i} B_{j}^{*}-\frac{1}{2}(\varepsilon_{0}|\mathbf{E}|^{2}+\frac{1}{\mu_{0}}|\mathbf{B}|^{2}) \delta_{i j}\right],
\label{eqn:6}
\end{equation}
where ``*" denotes the complex conjugate and $\delta_{i j}$ is the Kronecker delta.

For an intuitive understanding of the force and torque, we can apply the multipole expansion method. Adopting the source representation and using Einstein summation convention, the optical force and torque can be decomposed into the contributions of electromagnetic multipoles as \cite{Chen2011, nieto2015optical,Mobini2018}
\begin{equation}
\begin{aligned}
&\left(\mathbf{F}_{\text {inc}}\right)_{i}=\frac{1}{2} \operatorname{Re}\left[p_{j} \nabla_{i} E_{j}^{*}\right]+\frac{1}{2} \operatorname{Re}\left[m_{j} \nabla_{i} B_{j}^{*}\right]+\\
&\frac{1}{12} \operatorname{Re}\left[Q_{j k}^{e} \nabla_{i} \nabla_{k} E_{j}^{*}\right]+\frac{1}{12} \operatorname{Re}\left[Q_{j k}^{m} \nabla_{i} \nabla_{k} B_{j}^{*}\right] \\
\end{aligned}
\label{eqn:7}
\end{equation}

\begin{equation}
\begin{aligned}
&\left(\mathbf{F}_{\text {int}}\right)_{i}=-\frac{k^{4}}{12 \pi \varepsilon_{0} c} \operatorname{Re}\left[\varepsilon_{i j k} p_{j} m_{k}^{*}\right]
-\frac{k^{5}}{120 \pi \varepsilon_{0}} \operatorname{Im}\left[Q_{i j}^{e} p_{j}^{*}\right]\\
&-\frac{k^{5}}{120 \pi \varepsilon_{0}} \operatorname{Im}\left[Q_{i j}^{m} m_{j}^{*}\right]
-\frac{k^{6}}{9 \times 240 \pi \varepsilon_{0} c} \operatorname{Re}\left[\varepsilon_{i j k} Q_{l j}^{e} Q_{l k}^{m}\right]
\end{aligned}
\label{eqn:8}
\end{equation}

\begin{equation}
\left(\mathbf{G}_{\text {inc}}\right)_{i}=\frac{1}{2} \operatorname{Re}\left[\varepsilon_{i j k} p_{j} E_{k}^{*}\right]+\frac{1}{2} \operatorname{Re}\left[\varepsilon_{i j k} m_{j} B_{k}^{*}\right],
\label{eqn:9}
\end{equation}

\begin{equation}
\left(\mathbf{G}_\text{int}\right)_{i}=\frac{k^{3}}{12 \pi \varepsilon_{0}} \operatorname{Im}\left[\varepsilon_{i j k} p_{j} p_{k}^{*}\right]+\frac{\mu_{0} k^{3}}{12 \pi} \operatorname{Im}\left[\varepsilon_{i j k} m_{j} m_{k}^{*}\right].
\label{eqn:10}
\end{equation}
Here, $F_\text{inc}$, $F_\text{int}$, $G_\text{inc}$, and $G_\text{int}$ are the incident force, interference force, incident torque, and interference torque, respectively (they are also referred to as the interception force, recoil force, extinction torque, and recoil torque in the literature, respectively); $p_{j}$, $m_{j}$, $Q_{ij}^e$,  and $Q_{ij}^m$ are the Cartesian components of the electric dipole, magnetic dipole, electric quadrupole, and magnetic quadrupole, respectively. The incident force and incident torque emerge from the coupling between the incident fields and the induced multipoles. The interference force and interference torque are attributed to the coupling among the multipoles. The electromagnetic multipoles can be evaluated as  \cite{nieto2015optical,jiang2015universal,Alaee2018}


\begin{equation}
\begin{aligned}
& p_{i}=\\&-\frac{1}{i \omega} \left\{ \int d\tau J_{i} j_{0}(k r)+
\frac{k^{2}}{2} \int d\tau\left[3(\mathbf{r} \cdot \mathbf{J}) r_{\alpha}-r^{2} J_{i}\right] \frac{j_{2}(k r)}{(k r)^{2}} \right\}
\end{aligned}
\label{eqn:11}
\end{equation}

\begin{equation}
m_{i}=\frac{3}{2} \int d\tau(\mathbf{r} \times \mathbf{J})_{i} \frac{j_{1}(k r)}{k r}
\label{eqn:12}
\end{equation}

\begin{equation}
\begin{aligned}
&Q_{\alpha \beta}^{e}=-\frac{3}{i \omega} \left\{ \int d\tau [3\left(r\odot J)-2(\mathbf{r} \cdot \mathbf{J}) \delta_{\alpha \beta}\right] \frac{j_{1}(k r)}{k r}+ \right. \\
&\left. 2k^{2} \int d\tau\left[5 r_{\alpha} r_{\beta}(\mathbf{r} \cdot \mathbf{J})-(r\odot J)r^{2}-r^{2}(\mathbf{r} \cdot \mathbf{J}) \delta_{\alpha \beta}\right] \frac{j_{3}(k r)}{(k r)^{3}}\right\}
\end{aligned}
\label{eqn:13}
\end{equation}

\begin{equation}
Q_{\alpha \beta}^{m}=15 \int d\tau \ [r_{\alpha}(\mathbf{r} \times \mathbf{J})_{\beta}+r_{\beta}(\mathbf{r} \times \mathbf{J})_{\alpha} ] \frac{j_{2}(k r)}{(k r)^{2}}
\label{eqn:14}
\end{equation}
where $J_i$ is the Cartesian components of induced electric current density in the sphere, $k$ is the wave number, $r\odot J=r_{\beta} J_{\alpha}+r_{\alpha} J_{\beta}$, and $j_{1,2,3} (kr)$ are the spherical Bessel functions. We note that the electric dipole $p_i$ already contains the contribution of the toroidal dipole \cite{Savinov2014,Zhang2015,fernandez2017dynamic}.


The optical force and torque are closely related to the scattering properties of the spinning sphere, which can be characterized by the normalized scattering cross section 
\begin{equation}
C_{\mathrm{sca}}=\frac{Z_{0}}{\mu_{0}\left|E_{0}\right|^{2}} \oint_{S} \mathbf{S} \cdot \hat{\mathbf{n}} d a.
\label{eqn:15}
\end{equation}
Here, $Z_0$ is the impedance of the free space; $E_0$ is the amplitude of the incident electric field; $\bold{S}$ is the time-averaged Poynting vector of the scattering fields. The scattering cross section can also be decomposed into the contributions of multipoles \cite{Alaee2018}: 


\begin{equation}
\begin{aligned}
&C_{\mathrm{sca}}=\\ & \frac{k^{4}}{6 \pi \varepsilon_{0}^{2}\left|E_{0}\right|^{2}}\left[\left|p_{\alpha}\right|^{2}+\left|\frac{p_{\alpha}}{c}\right|^{2}+\frac{1}{120}\left(\left|k Q_{\alpha \beta}^{e}\right|^{2}+\left|\frac{k Q_{\alpha \beta}^{m}}{c}\right|^{2}\right)\right].
\end{aligned}
\label{eqn:16}
\end{equation}

\section{\label{sec: III. MST,SourceRep} RESULTS AND DISCUSSION }
We consider a silicon sphere with a radius of $r = 200$ nm under the illumination of an electromagnetic plane wave. The sphere is lossless with relative permittivity $\varepsilon_r=11.9$ and relative permeability $\mu_r=1$. The sphere spins about the $z$-axis with the angular velocity $\mathbf{\Omega } =\varOmega \uvec{z}$. The plane wave propagates in the $+z$ or $+x$ direction, corresponding to the parallel and perpendicular incidences, respectively, with respect to the spinning axis of the sphere. In what follows, we will focus on the force and torque in two cases: the sphere upon parallel incidence and the sphere upon the perpendicular incidence. The system in the former case has the cylindrical symmetry about the spinning axis, while this symmetry is broken in the latter case. All the numerical results are obtained by full-wave simulations using COMSOL Multiphysics.

\subsection{\label{sec: A} Optical force and torque under parallel incidence}
\begin{figure}
    \centering
    \includegraphics[width=1\linewidth]{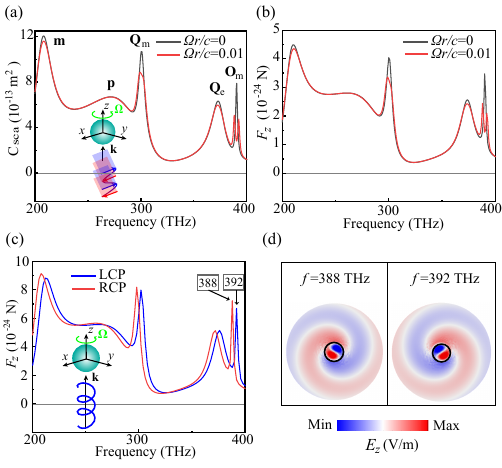}
    \caption{(a) Scattering cross section of a stationary ($\varOmega=0$) and a spinning ($\varOmega r/c=0.01$) sphere illuminated by a linearly polarized light. The inset shows the schematic of the considered system, where a sphere spinning around $z$-axis is illuminated by a linearly polarized light. (b) Optical force on the stationary and spinning spheres induced by a linearly polarized light. (c) Optical force on the spinning sphere induced by the LCP and RCP lights. The inset shows the schematic of the system. (d) The electric field $E_z$ at $z$=0 plane at the two resonances labeled by red and blue dots in (c).}
    \label{fig:1}
\end{figure}

We consider the spinning sphere illuminated by a linearly polarized plane wave $\bold{E}=\hat{x}e^{ikz-i\omega t}$, as shown by the inset in Fig. \ref{fig:1}(a). \textcolor{black}{The spinning sphere has the cylindrical symmetry with respect to $z$ axis as indicated by Eq. (3) (the constitute relation has no azimuthal dependence).} Thus, there is no lateral force induced on the sphere, and only forward scattering force can emerge. In addition, the cylindrical symmetry leads to vanished coupling among different azimuthal modes \cite{Waterman1971,Nieminen2004}. Consequently, the total optical angular momentum is conserved, and the plane wave cannot exert optical torque on the sphere via scattering. For a lossless sphere spinning about $z$ axis, there is no optical torque in $z$ direction, regardless of the incident wave settings. Despite the absence of lateral force and optical torque, this configuration provides an excellent opportunity to investigate the influence of the Sagnac effect on the optical force. 


We conduct full-wave simulation of the system and compute the scattering cross section of the sphere. The results are shown in Fig. \ref{fig:1}(a) for the stationary sphere with $\varOmega=0$ (corresponding to the black line) and for the sphere spinning at the normalized velocity $\varOmega r/c=0.01$ (corresponding to the red line). In the stationary case, five peaks appear in the spectrum, which are labeled as $\mathbf{m},\mathbf{p},\mathbf{Q}_m,\mathbf{Q}_e,\mathbf{O}_m$ and correspond to the resonances of magnetic dipole, electric dipole, magnetic quadrupole, electric quadrupole, and magnetic octupole, respectively. For the spinning sphere, we notice that the resonance peak of the magnetic octupole splits into two peaks. This phenomenon corresponds to the Sagnac effect, and the emergence of the two peaks is attributed to the frequency splitting of the chiral octupole modes in the sphere. Without spinning motion, the sphere supports two degenerate and orthogonal chiral magnetic octupole modes in the circular basis $(\hat{x}\pm i\hat{y})/\sqrt{2}$; one rotates in clockwise direction while the other rotates in counterclockwise direction. The degeneracy of the two modes is broken by the spinning motion of the sphere, and the two chiral octupole modes have different eigen frequencies. Under the linearly polarized plane wave incidence, both chiral octupole modes are excited, giving rise to the two peaks in the scattering spectrum. Notably, this frequency splitting phenomenon also happens to other multipole modes, including the dipoles and quadrupoles, which is not visible in the scattering spectrum in Fig. \ref{fig:1}(a) due to the low quality factors of the modes. The frequency splitting between two chiral multipole modes generally increases with the spinning speed and multipole order. For a two-dimensional infinitely-long cylinder under spinning motion, where TE and TM polarizations are decoupled, a closed-form analytic expression can be derived for the frequency splitting \cite{Shi2021,Yang2022}. Such an analytical expression without approximation cannot be obtained for the 3D sphere considered here due to the complicated coupling of the electric and magnetic fields. 

We apply the Maxwell stress tensor approach to calculate the optical force induced on the sphere. The results are shown in Fig. \ref{fig:1}(b) for the stationary sphere (denoted by the black line) and spinning sphere (denoted by the red line). The force is along the propagation direction ($+z$) of the incident plane wave. We notice that the force has a similar spectrum as the scattering cross section. In particular, the force also exhibits a frequency splitting due to the Sagnac effect. This is expected since the optical force is entirely attributed to the light scattering by the sphere. 

We also calculate the optical force when the spinning sphere is illuminated by a circularly polarized plane wave $\bold{E}=(\hat{x}\pm i\hat{y})e^{ikz-i\omega t}$. Figure \ref{fig:1}(c) shows the optical forces when the incident light is left-handed circularly polarized (LCP) and right-handed circularly polarized (RCP), denoted by the blue and red lines, respectively. As seen, the force spectra are different for the LCP and RCP lights. The spectra clearly exhibit a frequency shift for all the considered multipoles. The resonance peaks locate at different frequencies for the LCP and RCP excitations. This is because that the LCP and RCP plane waves selectively excite opposite chiral multipole modes, as shown in Fig. \ref{fig:1}(d) for the magnetic octupole case, where electric fields $E_z$ of the two modes clearly have opposite chirality. The selective excitation is attributed to the match of the spin angular momentum of the incident wave and the chiral multipoles \cite{Yang2022}.


\subsection{\label{sec: B} Optical force and torque under perpendicular incidence}
\begin{figure}
    \centering
    \includegraphics[width=1\linewidth]{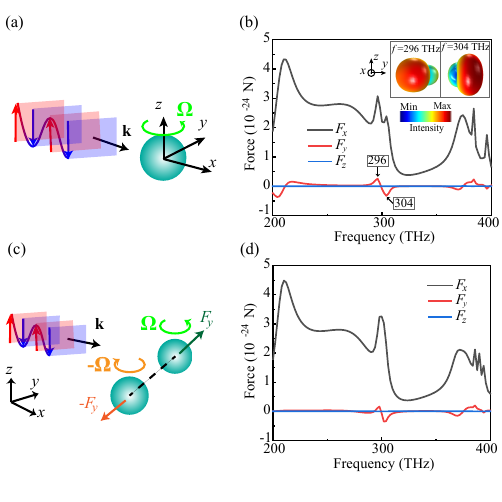}
    \caption{(a) Spinning sphere under the incidence of a linearly polarized light. (b) Optical force induced on the spinning sphere. The inset shows the Far-field scattering intensity distribution at the frequencies 296 THz and 304 THz. (c) A plane wave is able to drag spheres spinning in opposite directions away from each other. (d) Optical force on the spinning sphere induced by the y-polarized plane wave.}
    \label{fig:2}
\end{figure}

We consider the spinning sphere with $\varOmega r/c=0.01$ illuminated by a plane wave propagating in $x$ direction, as shown in Fig. \ref{fig:2}(a). In this case, the mirror symmetry with respect to the $xoz$ plane is broken, which results in asymmetric scattering and thus a lateral optical force. Figure \ref{fig:2}(b) and \ref{fig:2}(d) show the numerically calculated optical forces when the plane wave is linearly polarized in $z$ direction and $y$ direction, respectively. We notice that in both cases the optical force is dominated by the forward component $F_x$. The force spectra are different from Fig. \ref{fig:1}(b) due to the broken mirror symmetry. Interestingly, a lateral force perpendicular to the rotating axis appears in both cases. Since the incident plane wave carries no linear momentum in the lateral direction, this lateral force is entirely attributed to the scattered momentum of light. To confirm this, as shown in the inset of Fig. \ref{fig:2}(b), we plot the far-field scattering intensity at 296 THz and 304 THz under the $z$-polarized incidence, corresponding to local maximum and minimum of the lateral force. It is clear that the incident wave is predominantly scattered to the $-y$ direction at 296 THz, resulting in a lateral force along the $+y$ direction. In contrast, the incident wave is mainly scattered to $+y$ direction at 304 THz, giving rise to a lateral force in $-y$ direction. In addition, Fig. \ref{fig:2}(b) and (d) show that there is no lateral force in the $z$-direction, \textcolor{black}{which is attributed to the mirror symmetry of the spinning sphere with respect to the $xoy$ plane.} 

The above results demonstrate that a plane wave can induce a lateral optical force on a temporally chiral sphere, where the \textit{temporal chirality} originates from the dynamic rotation of the sphere and is equivalent to the effective bi-anisotropy. Importantly, the direction of the lateral force directly depends on the sign of the temporal chirality (i.e., spinning handedness) of the sphere. Under the incidence of a linearly polarized plane wave, the sphere spinning in opposite sense is subject to the lateral force pointing in opposite directions. Consequently, the lateral force is capable of sorting temporally chiral particles with opposite chirality, as schematically shown in Fig. \ref{fig:2}(c). This is in stark contrast to the lateral optical force on conventional chiral particles (e.g., helix particles) \cite{Wang2014}, which is attributed to the interaction between \textit{spatial chirality} and light fields.

\begin{figure}
    \centering
    \includegraphics[width=1\linewidth]{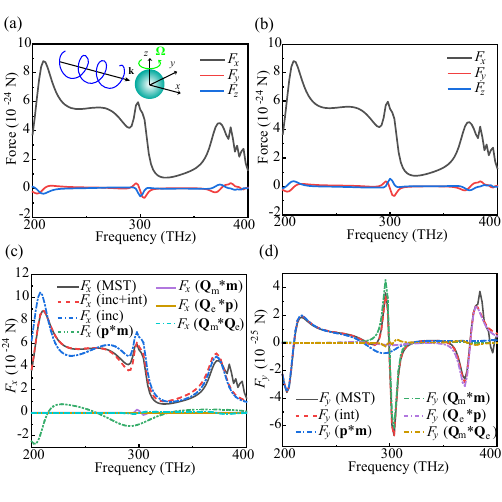}
    \caption{Optical force on the spinning sphere induced by the (a) LCP and (b) RCP light. The inset in (a) shows the system with a spinning sphere under the incidence of a circularly polarized light. Multipole expansion for the force components of (c) $F_x$ and (d) $F_y$ under the LCP excitation. The solid black lines denote the results obtained with Maxwell stress tensor approach.}
    \label{fig:3}
\end{figure}

Figures \ref{fig:3}(a) and \ref{fig:3}(b) show the optical forces induced on the spinning sphere with $\varOmega r/c=0.01$ by the incident LCP and RCP light, respectively. We notice that lateral forces emerge in both $y$ and $z$ directions in the two cases, in contrast to the cases of Fig. \ref{fig:2} where lateral force only appears in $y$ direction. This difference is attributed to the absence of mirror symmetry with respect to the $xoz$ and $xoy$ planes in the cases of Figs. \ref{fig:3}(a) and \ref{fig:3}(b). In addition, we see that the forward force $F_x$ (denoted by the black line) and the lateral force $F_y$ (denoted by the red line) have the same profile in both cases. Meanwhile, the lateral force $F_z$ (denoted by the blue line) has the same magnitude but opposite directions. These results can be understood with a simple symmetry analysis. Since a mirror operation of the LCP system with respect to the $xoy$ plane transforms it to the RCP system, the optical forces in both cases must satisfy the relationship: $ F_x^{\mathrm{LCP}}=F_x^{\mathrm{RCP}}$, $F_y^{\mathrm{LCP}}=F_y^{\mathrm{RCP}}$, and $F_z^{\mathrm{LCP}}=-F_z^{\mathrm{RCP}}$.

To further understand the origin of the optical forces, we apply multipole expansion method to decompose the forces into the contributions of electromagnetic multipoles. We consider the case of Fig. \ref{fig:3}(a) as an example. The decompositions of the forward force $F_x$ and lateral force $F_y$ are summarized in Figs. \ref{fig:3}(c) and \ref{fig:3}(d), respectively. The analysis for the lateral force $F_z$ is similar and not shown here. For $F_x$ in Fig. \ref{fig:3}(c), we decompose it into the incident force and the interference forces contributed by the multipoles up to the quadrupole. The black line denotes the total force $F_x$ obtained with the MST method. The dashed red line corresponds to the sum of the incidence force and interference forces, which agrees with the MST result well. We notice that the incident force (blue dash-dot line) is always positive, while the interference forces can be either positive or negative due to the different phases of the multipoles. 

According to Eq. (8), the time-averaged lateral force $F_y$ is induced by the interferences of the multipoles as
\begin{equation}
\begin{aligned}
F_y=&-\frac{k^{4}}{12 \pi \varepsilon_{0} c} \text{Re}[p_z m_x^*-p_x m_z^* ]-\frac{k^{5}}{12 \pi \varepsilon_{0} } \text{Im}[Q_{yj}^e p_j^* ]\\
&-\frac{k^{5}}{12 \pi \varepsilon_{0} c^2} \text{Im}[Q_{yj}^m m_j^* ]. 
\end{aligned}
\end{equation}
The contribution of each term is shown in Fig. \ref{fig:3}(d). The solid black line denotes the total force obtained with the MST method. The red dash line denotes the sum of the interference forces contributed by the multipoles (i.e., the sum of all the dashed-dot lines), which agrees well with the MST result. We notice that the lateral force is strongly enhanced at the resonance frequencies of the multipoles. Specifically, the strong lateral force at about 200 THz is induced by the interference of the electric dipole $\mathbf{p}$ and magnetic dipole $\mathbf{m}$; the strong lateral force at about 300 THz is mainly attributed to the interference of the magnetic dipole $\mathbf{m}$ and magnetic quadrupole $\mathbf{Q}_m$; the strong lateral force at about 380 THz is mainly attributed to the interference of electric dipole $\mathbf{p}$ and electric quadrupole $\mathbf{Q}_e$. Notably, the lateral force undergoes a sign change near resonance frequencies. This is because that at the resonant frequencies, one multipole undergoes a phase change of $180^{\circ}$  while the phase of the other multipole remains approximately unchanged due to off resonance. This mechanism enables flexible control of the lateral force by adjusting the frequency of the incident light.

\begin{figure}
    \centering
    \includegraphics[width=1\linewidth]{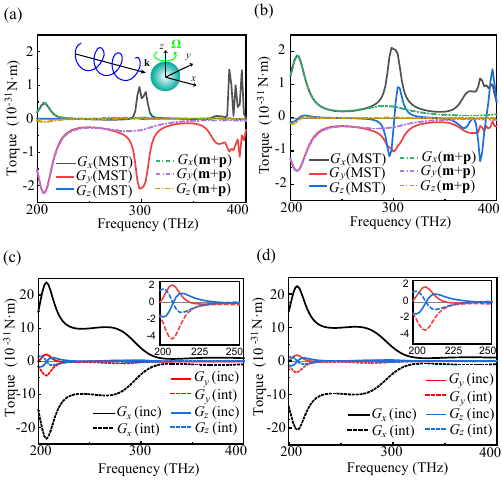}
    \caption{Optical torque induced by the LCP light on the (a) lossless (b) lossy spinning spheres. The dash lines denote the dipole contributions. The inset in (a) shows the system with a spinning sphere under the incidence of a circularly polarized light. Incident torque and interference torque for the (c) lossless and (d) lossy spheres. Only the contributions of the electric dipole $\mathbf{p}$ and magnetic dipole $\mathbf{m}$ are shown. The insets show the zoom-ins of (c) and (d) with the same unit.}
    \label{fig:4}
\end{figure}

Figure \ref{fig:4}(a) shows the optical torque induced on the spinning sphere by the LCP plane wave (the torque induced by the RCP plane wave can be easily obtained with a mirror operation). We notice that the torque is nonzero despite that the sphere is made of isotropic and lossless silicon. This is different from the optical torque induced by circularly polarized light on a stationary lossless silicon sphere, which is always zero because the total optical angular momentum remains unchanged in the scattering process. For the spinning sphere, the nonzero optical torque is attributed to the breaking of cylindrical symmetry by the spinning motion. As shown in Fig. \ref{fig:4}(a), the LCP light exerts a positive torque on the sphere in the $x$ direction (i.e., $G_x$) and a negative torque in the $y$ direction (i.e., $G_y$). The torques $G_x$ and $G_y$ are strongly enhanced at the resonance frequencies of the multipoles. In addition, the optical torque does not have a component in $z$ direction due to the cylindrical symmetry of the spinning sphere with respect to the $z$ axis. To understand the effect of absorption, we also calculate the optical torque on the spinning silicon sphere with absorption by setting the permittivity to be $\varepsilon_r=11.9+0.1i$ . The results are shown in Fig. \ref{fig:4}(b). We notice that the torque $G_x$ is enhanced while the torque $G_y$ is reduced. Interestingly, an optical torque component along $z$ direction also appears in the lossy case. We apply multipole expansions in Eq. (7) to understand the first resonance in both cases, considering only the contributions of electric dipole $\mathbf{p}$ and magnetic dipole $\mathbf{m}$. The results are denoted as the dashed lines in Figs. \ref{fig:4}(a) and \ref{fig:4}(b), which shows good consistency between the numerical and analytical results. Thus, the first resonance indeed is mainly attributed to the electric and magnetic dipoles. 

To understand the variations of the optical torque due to absorption, we further decompose the optical torque into the incident torque and interference torque. The results are shown in Figs. \ref{fig:4}(c) and \ref{fig:4}(d) for the cases without and with loss, respectively. When a circularly polarized light illuminates the stationary lossless sphere, the incident torque and interference torque cancel each other, resulting in zero net torque. The spinning motion of the sphere leads to a difference between the extinction and interference  torques, as shown in Fig. \ref{fig:4}(c), thus giving rise to nonzero $G_x$ and $G_y$. We note that $G_z$ remains zero as a result of the cylindrical symmetry. When loss is introduced into the spinning sphere, as shown in Fig. \ref{fig:4}(d), the magnitudes of the incident and interference torques decrease differently, and the net torque in $z$ direction is not zero anymore. Besides, the absorption has a stronger effect on the interference torque than on incident torque, leading to the increase of $G_x$ and $G_y$ (i.e. less negative for $G_y$). Finally, we note that a linearly polarized plane wave propagating in $x$ direction can also exert optical torque $G_x$ and $G_y$ on the sphere spinning around $z$ axis (except for the $y$- and $z$-polarized incidences that preserve the mirror symmetry about $xoy$ plane).


\section{\label{sec: IV. MST,SourceRep} CONCLUSION}
We investigate the optical force and torque on a spinning silicon sphere induced by linearly polarized and circularly polarized plane waves. We discovered several interesting phenomena originating from the symmetry breaking and effective bi-anisotropic property of the sphere, including the Sagnac effect, lateral optical force, and lateral optical torque. When the incident direction of the plane wave is parallel to the spinning axis of the sphere, the cylindrical symmetry about the spinning axis is preserved, and lateral force and optical torque are prohibited. In this case, we discuss the Sagnac effect on the optical force, which gives rise to two resonance peaks in the force spectrum. When the incident direction is not parallel to the spinning axis, the symmetry breaking leads to unusual lateral force and optical torque, which do not exist in the conventional stationary system involving the interaction between a plane wave and a lossless isotropic sphere. The emergence of these unusual optical force and torque are directly related to the spinning-induced bi-anisotropy of the sphere. By applying multipole expansion, we uncover how the electromagnetic multipoles induced in the sphere give rise to the optical force and torque. Our work contributes to the understanding of the optical force and torque in moving media. The results can find applications in realizing unconventional optical manipulation of small particles.

\section{\label{sec: VII. Acknowledgements}Acknowledgements}
The work described in this paper was supported by the National Natural Science Foundation of China (No. 12322416) and the Research Grants Council of the Hong Kong Special Administrative Region, China (Projects No. CityU 11301820 and No. AoE/P-502/20). 

\bibliography{mendeley}
\end{document}